\documentclass{article}
\usepackage{gsirep}
\usepackage{graphicx}
\title{Density Perturbations in Heavy-Ion Collisions below the
Critical Point}
\author{A.~Dumitru, K.~Paech and H.~St\"ocker}
\affil{Institut f\"ur Theoretische Physik, Universit\"at Frankfurt a.M., Germany}
\date{}
\begin{document}
\maketitle
\noindent 

Universality arguments suggest that the chiral phase transition for 
two massless quark flavors is second-order
at baryon-chemical potential $\mu_B=0$~\cite{PW},
which then becomes a crossover for small quark masses.
On the other hand, a first-order phase transition is predicted by a variety
of low-energy effective theories for small
temperature $T$ and large $\mu_B$~\cite{1stO}.
Hence, the first-order phase transition line in the $(\mu_B,T)$ plane must end 
in a second-order critical point~\cite{Stephanov:1999zu}.
For $2+1$ quark flavors the critical point has been located at
$T=160$~MeV and $\mu_B= 725$~MeV~\cite{Fodor:2001pe}.
However, a reliable extrapolation to the continuum limit and to physical
pion mass has not been attempted so far.

There is an ongoing experimental effort to detect that critical point
in heavy-ion collisions at high energies. It is hoped that by varying the
beam energy, for example, one can ``switch'' between the regimes of first-order
transition and cross over, respectively (higher energies correspond to larger
entropy per baryon or $T/\mu_B$).

\begin{figure}[ht]
\includegraphics[width=8cm]{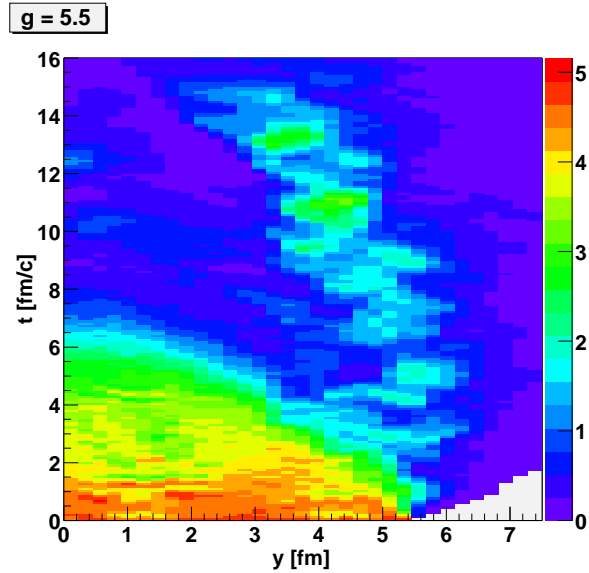}
\vspace*{-.3cm}
\caption{Fluid energy density distribution in space-time for a first-order
chiral phase transition~\protect\cite{Paech:2003fe}.}
\label{edens_chhyd}
\vspace*{-.4cm}
\end{figure}

To investigate collective dynamics in the vicinity of the critical endpoint
we introduce a model for the real-time evolution of a relativistic
fluid of quarks coupled to non-equilibrium dynamics of the long wavelength
(classical) modes of the chiral condensate~\cite{Paech:2003fe}:
\begin{equation}
\partial_\mu\partial^\mu \phi + \partial V_{\rm eff}/\partial\phi 
=0~,~\partial_\mu \left( T_{\rm fl}^{\mu\nu} + T_\phi^{\mu\nu}\right)=0.
\end{equation}
Here, $T_{\rm fl}^{\mu\nu}$ is the energy-momentum tensor of the fluid,
$T_\phi^{\mu\nu}$ that of the classical modes of the chiral condensate, and
$V_{\rm eff}$ is the effective potential obtained by integrating out the
thermalized degrees of freedom.
We focus first on energy-density inhomogeneities for vanishing baryon
density (the nature of the transition is then determined by the effective
quark-field coupling rather than the baryon-chemical
potential~\cite{Paech:2003fe,ove}).
We allow for ``primordial'' Gaussian fluctuations of the condensate $\phi$
on length scales
$\sim 1$~fm on top of a smoothly varying mean field. If propagated
through a first-order chiral phase transition these fluctuations give rise to
a rather inhomogeneous (energy-) density distribution
as seen in Fig.~\ref{edens_chhyd}. Such effects were previously studied
in the context of the QCD transition in the early universe, where
inhomogeneities of the entropy (or baryon to photon ratio) might affect
BBN~\cite{BBN}. However, in the cross-over regime we find much smaller
amplitudes of density perturbations~\cite{Paech:2003fe}.

In heavy-ion collisions the scale of the density perturbations is too small
for them to be resolved in rapidity space. This would require a resolution
$\Delta y<1$, which is about
the thermal width of the local particle momentum distributions.
However, observable consequences of large density inhomogeneities created in
a first-order transition at beam energies below the critical endpoint
may still exist. 
(Inhomogeneities from fluctuations of particle production in the primary
nucleon-nucleon collisions~\cite{Bleicher:wd}
should be largely washed out until decoupling.)
For example, fluctuations of the energy-momentum tensor of matter in
coordinate space are uncorrelated to the reaction plane and
should therefore reduce out-of-plane collective flow ($v_2/\langle
p_t\rangle$) as compared to equilibrium hydrodynamics~\cite{Paech:2003fe}.
Moreover, by analogy to BBN, perturbations of the
entropy per baryon $s/\rho_B$ should affect abundances of rare hadrons:
$\bar{B}/B$, $\bar\Lambda/\bar{p}$~\cite{Back:2001ai}
and $K^+/\pi^+$~\cite{NA49} are larger than for a
homogeneous system with the same total volume, baryon number and entropy.

\end{document}